\documentclass[prb,twocolumn,showpacs]{revtex4}   

\usepackage{graphicx,color}

\begin{document}

\title
{
Quantum Monte Carlo study of nonequilibrium transport through 
a quantum dot coupled to normal and superconducting leads
}

\author
{
Akihisa Koga
}

\affiliation
{
Department of Physics, Tokyo Institute of Technology, Meguro, Tokyo 152-8551, Japan
}

\date{\today}

\begin{abstract}
We investigate the nonequilibrium phenomena through the quantum dot coupled to
the normal and superconducting leads 
using a weak-coupling continuous-time Monte Carlo method.
Calculating the time evolution of particle number, double occupancy,
and pairing correlation at the quantum dot,
we discuss how the system approaches the steady state.
We also deduce the steady current through the quantum dot 
beyond the linear response region.
It is clarified that the interaction decreases the current in the Kondo-singlet 
dominant region.
On the other hand, when the quantum dot is tightly coupled to 
the superconducting lead, the current is increased by the introduction of 
the Coulomb interaction,
which originates from the competition between the Kondo and proximity effects.
Transient currents induced by the interaction quench are also addressed.
\end{abstract}

\pacs{Valid PACS appear here}%

\maketitle

\section{Introduction}\label{1}
Recently electron transport through nanofabrications 
has attracted much interest.
One of the simplest systems is a quantum dot 
with discrete energy levels,\cite{dot}
which gives us a stage to study fundamental quantum physics.
When the quantum dot is contacted with the normal leads, 
electron correlations play a crucial role 
in understanding its transport at low temperatures
where the Coulomb blockade and Kondo effect appear.
On the other hand, when superconducting leads 
are connected to the quantum dot,\cite{dotS}
the proximity-induced on-dot pairing becomes important,
in addition to electron correlations.
However, multiple Andreev reflections should dominate the system 
and it is difficult to study the interplay 
between the superconducting and 
electron correlations systematically.

The quantum dot system coupled to the normal and 
superconducting leads is one of the appropriate systems to
study how the transport properties are affected by the competition between 
the Kondo and proximity-induced on-dot pairing effects.
In fact, the system has experimentally been examined,
\cite{Graber,Hofstetter,Herrmann} and 
the Kondo-enhanced Andreev transport has recently been observed in 
the InAs quantum dot.\cite{Deacon,Deacon2}
Theoretical study has been done by many groups,
\cite{slave1,slave2,Domanski,nca,Cuevas,Yamada,Tanaka,Baranski,Koerting}
and some interesting transport properties have successfully been explained.
However, it is nontrivial how the techniques are applicable
in the strong coupling and high voltage region.
This may be important to understand the experimental results quantitatively
since the linear response region is narrow in the quantum dot system.
Therefore, the unbiased and robust method for the nonequilibrium phenomena
is desired.

To this end, we make use of the continuous-time quantum Monte Carlo (CTQMC) 
method~\cite{CTQMC} based on the Keldysh formalism.~\cite{WernerOka,Werner}
Here we extend the CTQMC method in the continuous-time auxiliary field 
(CTAUX) formulation~\cite{Gull} 
to treat the superconducting state in the Nambu formalism.
Calculating the particle number, double occupancy, pairing correlations and 
current through the quantum dot,
we study the nonequilibrium phenomena.
We also discuss the competition between the Kondo and proximity effects on
the steady-state in the quantum dot
coupled to the normal and superconducting leads. 

The paper is organized as follows.
In Sec. \ref{2}, we introduce the model Hamiltonian for the quantum dot coupled to
the normal and superconducting leads. 
The CTQMC algorithm in the Nambu formalism is explained in Sec. \ref{3}.
In Sec. \ref{4}, we discuss the nonequilibrium phenomena in the quantum dot system.
A brief summary is given in Sec. \ref{5}.

\section{Model}\label{2}
We consider the electron transport through the quantum dot coupled to 
the normal and superconducting leads,
which are labeled by $\alpha=N$ and $S$.
For simplicity, we use a single level quantum dot 
with the Coulomb interaction. 
The system should be described by the 
following Anderson impurity Hamiltonian as
\begin{eqnarray}
H&=&H_{bath}+H_{hyb}+H_{dot},\label{Hami}\\
H_{bath}&=&\sum_{k\alpha\sigma}\left(\epsilon_{k\alpha}-\mu_\alpha\right)
c_{k\alpha\sigma}^\dag c_{k\alpha\sigma}\nonumber\\
&+&
\sum_k\left(\Delta_S c_{-kS\downarrow}^\dag c_{kS\uparrow}^\dag+\Delta_S^* c_{kS\uparrow} c_{-kS\downarrow}\right),\\
H_{hyb}&=&\sum_{k\alpha\sigma}\left(
V_{k\alpha} c_{k\alpha\sigma}^\dag d_\sigma+
V_{k\alpha}^* d_\sigma^\dag c_{k\alpha\sigma} \right),\\
H_{dot}&=&H_{dot}^0+H',\\
H_{dot}^0&=&
\sum_\sigma\left(\epsilon_d+\frac{U}{2}\right)n_\sigma,\\
H'&=&
U\left( n_\uparrow n_\downarrow 
-\frac{1}{2}\sum_\sigma n_\sigma\right),\label{U}
\end{eqnarray}
where $c_{k\alpha\sigma} (c_{k\alpha\sigma}^\dag)$ 
is the annihilation (creation) 
operator of an electron with wave vector $k$ and 
spin $\sigma(=\uparrow, \downarrow)$ 
in the $\alpha$th lead.
$d_{\sigma} (d_{\sigma}^\dag)$ is 
the annihilation (creation) 
operator of an electron at the quantum dot and $n_\sigma=d_\sigma^\dag d_\sigma$.
$\epsilon_{k\alpha}$ is the dispersion relation of the $\alpha$th lead and 
$V_{k\alpha}$ is the hybridization 
between the $\alpha$th lead and the quantum dot. 
$\epsilon_d$ and $U$ is the energy level and the Coulomb interaction
at the quantum dot.
To discuss the nonequilibrium state in the system, we set 
the chemical potential in each lead as $\mu_N=V$ and $\mu_S=0$, where
$V$ is the bias voltage.
In our paper, we focus on the particle-hole symmetric system
with $\epsilon_d+U/2=0$ and the superconducting lead is assumed to be described 
by the BCS theory with an isotropic gap $\Delta_S=\Delta(>0)$.
We consider a sufficiently wide bandwidth in both leads,
where the coupling strength 
$\Gamma_\alpha(\omega)=\pi\sum_k\left|V_{k\alpha}\right|^2
\delta(\omega-\epsilon_{k\alpha})$
becomes constant.

When no bias voltage is applied to the system,
ground state properties depend on the ratio $\Gamma_S/\Gamma_N$.
In the case $\Gamma_S/\Gamma_N\ll 1$ and $U\neq 0$, 
conduction electrons in the normal lead screen the local spin 
at the quantum dot and  the Kondo-singlet dominant state is realized. 
On the other hand, the singlet Cooper pairs 
are realized at a quantum dot due to the proximity effect 
when $\Gamma_S/\Gamma_N\gg 1$.
It is known that when the ratio is changed, 
the crossover, in general, occurs between these two singlet states
and the first-order transition occurs in the limit $\Gamma_N=0$.
\cite{Yamada,Koerting,Satori}

This crossover affects nonequilibrium properties in the quantum dot system. 
It has been reported how the zero bias conductance\cite{Cuevas,Tanaka}
and the current-voltage characteristics\cite{Koerting} are
affected in the vicinity of the crossover.
Furthermore, a detailed structure in the differential conductance has been 
discussed in the nonlinear response region 
by means of the modified perturbation theory (MPT),\cite{Yamada} 
which is based on an interpolation scheme between 
the weak-coupling limit $(U\rightarrow 0)$ and 
the superconducting atomic limit $(\Gamma_N=0$, $\Delta\rightarrow\infty$).
Although the reliable results have been obtained in the cases 
$U/\Gamma_N \ll 1$ and $\Gamma_S/\Gamma_N\gg 1$,
it is unclear whether  
the nonlinear response in the strong coupling region
is quantitatively described or not.

In this paper, we make use of the CTQMC method. 
In the method, Monte Carlo samplings of collections of diagrams
are performed in continuous time and 
thereby the Trotter error,
which originates from the Suzuki-Trotter decomposition, is avoided.
This method has successfully been applied to many equilibrium systems.
Recently, the CTQMC method based on the Keldysh formalism has been formulated,
where the nonequilibrium phenomena in the quantum dot coupled to normal leads
have quantitatively been studied.\cite{WernerOka,Werner}
In the following, we extend the CTQMC method to deal 
 with the superconducting state in the Nambu formalism.

\section{Continuous-Time Quantum Monte Carlo Simulations in the Nambu Formalism}
\label{3}

\begin{figure}[htb]
\begin{center}
\includegraphics[width=6cm]{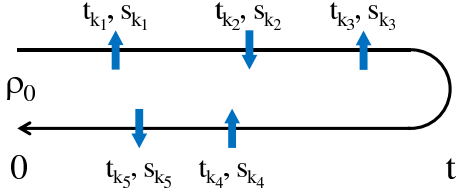}
\end{center}       
\caption{
Illustration of the Keldysh contour for the CTQMC method.
Arrows represent auxiliary Ising spins for 
a certain Monte Carlo configuration corresponding
to the perturbation order $l=3$ and $m=2$ $(n=5)$.
}
\label{contour}
\end{figure}

In this section, we explain the CTQMC method 
based on the Keldysh formalism,\cite{WernerOka,Werner}
and extend it to treat the superconducting state.
Here, we consider a weak-coupling version of the CTQMC approach.
Since the interaction is considered as a perturbation,
we can examine the time evolution of the system after the interaction quench.
We start with the following identity
\begin{eqnarray}
1&=&{\rm Tr}\left[\rho_0 e^{it\left(H_0+H'-K/t\right)}
e^{-it\left(H_0+H'-K/t\right)}\right],
\end{eqnarray}
where $\rho_0 (= e^{-\beta H_0}/{\rm Tr} e^{-\beta H_0})$ is 
the initial density matrix for $H_0(=H-H')$ and
$K$ is some nonzero constant.
It is then given by
\begin{eqnarray}
1&=&{\rm Tr}\Big\{\rho_0 
\tilde{T}\left[\exp\Big\{i\int_0^t d\tilde{t} \left(H'(\tilde{t})-K/t\right)
\Big\}\right]
e^{itH_0}\nonumber\\
&\times&
e^{-itH_0}
T\left[\exp\Big\{-i\int_0^t dt \left(H'(t)-K/t\right)\Big\}\right]\Big\},\nonumber
\end{eqnarray}
where $O(t)=e^{itH_0}Oe^{-itH_0}$ is the interaction representation 
of the operator $O$ and $T (\tilde{T})$ is 
the time-ordering (antitime-ordering) operator.
Expanding the exponentials into a power series, we obtain 
\begin{eqnarray}
1&=&{\rm Tr}\Big[\rho_0 \sum_l(-\frac{iK}{t})^l\int_{0}^t d\tilde{t}_1\cdots
\int_{\tilde{t}_{l-1}}^t d\tilde{t}_l e^{i\tilde{t}_1H_0}\nonumber\\
&\times& (1-\frac{t}{K}H')\cdots e^{i(\tilde{t}_l - \tilde{t}_{l-1}) H_0}
(1-\frac{t}{K}H') e^{i(t - \tilde{t}_{l}) H_0}\nonumber\\
&\times&\sum_m (\frac{iK}{t})^m
\int_{0}^t dt_1\cdots\int_{t_{m-1}}^t dt_m
e^{-i(t - t_m)H_0}\nonumber\\
&\times&(1-\frac{t}{K}H')\cdots e^{-i(t_2-t_1)H_0}(1-\frac{t}{K}H')e^{-it_1H_0}\Big]
\nonumber\\
&=&\frac{1}{{\rm Tr}\left[e^{-\beta H_0}\right]}\sum_{lm}(-i)^l i^m\left(\frac{K}{2t}\right)^{l+m}\sum_{\{\tilde{s}\}\{s\}}
\nonumber\\
&\times&
\int_0^t d\tilde{t}_1\cdots\int_{\tilde{t}_{l-1}}^t d\tilde{t}_l
\int_0^t dt_1\cdots\int_{t_{m-1}}^t dt_m\nonumber\\
&\times&{\rm Tr}\Big[
e^{-\beta H_0}e^{i{\tilde t}_1 H_0}e^{\gamma \tilde{s}_1
(n_\uparrow-n_\downarrow)}\cdots 
e^{i({\tilde t}_l-{\tilde t}_{l-1}) H_0}\nonumber\\
&\times&e^{\gamma \tilde{s}_l(n_\uparrow-n_\downarrow)}
e^{-i({\tilde t}_l-t_m) H_0}e^{\gamma s_m(n_\uparrow-n_\downarrow)}
\cdots\nonumber\\
&\times&e^{-i(t_2-t_1) H_0}e^{\gamma s_1(n_\uparrow-n_\downarrow)}
e^{-it_1H_0}\Big],
\end{eqnarray}
where we have used the following equation as
\begin{eqnarray}
1-\frac{tU}{K}\left(n_\uparrow n_\downarrow-\frac{1}{2}
\sum_\sigma n_\sigma\right)
=\frac{1}{2}\sum_{s=\pm 1}e^{\gamma s (n_\uparrow-n_\downarrow)}
\end{eqnarray}
with $\gamma = \cosh^{-1}(1+tU/2K)$. The introduction of 
the Ising variable $s$ in $H'$ allows us to perform Monte Carlo simulations.
An $(l+m)$th order configuration $c=\{s_{k_1}, s_{k_2}, \cdots, s_{k_n};
t_{k_1}, t_{k_2}, \cdots, t_{k_n} \}$ is represented by the auxiliary spins 
$s_{k_1}, s_{k_2}, \cdots, s_{k_n}$ at the Keldysh times 
$t_{k_1}, t_{k_2}, \cdots, t_{k_n}$ along the Keldysh contour,
where the $l (m)$ denotes the number of spins 
on the forward (backward) contour (see Fig. \ref{contour})
and $n=l+m$. 
Its weight $w_c$ is then given as
\begin{eqnarray}
w_c &=& (-i)^l i^m\left(\frac{Kdt}{2t}\right)^n 
\det \left[{\hat N}^{(n)}\right]^{-1},\label{weight}
\end{eqnarray}
where ${\hat N}$ is an $n\times n$ matrix
and its element consists of a $2\times 2$ matrix: 
\begin{equation}
\begin{array}{rcl}
\left[{\hat N}^{(n)}\right]^{-1}&=&{\hat \Gamma}^{(n)}-{\hat g}^{(n)}
\left({\hat \Gamma}^{(n)}-{\hat I}^{(n)}\right),\\
{\hat I}^{(n)}_{ij}&=&\delta_{ij}{\hat\sigma}_0,\\
{\hat \Gamma}^{(n)}_{ij}&=&\delta_{ij}
\exp\left({\gamma s_{k_i}{\hat\sigma}_z}\right),\\
{\hat g}^{(n)}_{ij}&=&{\hat \sigma}_z{\hat G}_0(t_{k_i}, t_{k_j}),
\end{array}
\end{equation}
with $i,j=1,2,\cdots, n$. The Green's function is explicitly given by
\begin{equation}
\begin{array}{rcl}
{\hat G}_0(t'_k,t''_k)&=&\left\{
\begin{array}{ll}
{\hat G}_0^<(t',t''),& t'_k<t''_k\\
{\hat G}_0^>(t',t''),& t'_k\ge t''_k
\end{array}
\right.
\end{array},
\end{equation}
where 
the times $t'$ and $t''$ correspond to the Keldysh times $t_k'$ and $t_k''$.
The lesser and greater Green's functions are defined by a $2\times 2$ matrix as
\begin{equation}
\begin{array}{rcl}
{\hat G}_0^<(t',t'')&=&i\left(
\begin{array}{cc}
\langle d_\uparrow^\dag(t'')d_\uparrow(t')\rangle&
\langle d_\downarrow(t'')d_\uparrow(t')\rangle\\
\langle d_\uparrow^\dag(t'')d_\downarrow^\dag(t')\rangle&
\langle d_\downarrow(t'')d_\downarrow^\dag(t')\rangle
\end{array}
\right),\\
{\hat G}_0^>(t',t'')&=&-i\left(
\begin{array}{cc}
\langle d_\uparrow(t')d_\uparrow^\dag(t'')\rangle&
\langle d_\uparrow(t')d_\downarrow(t'')\rangle\\
\langle d_\downarrow^\dag(t')d_\uparrow^\dag(t'')\rangle&
\langle d_\downarrow^\dag(t')d_\downarrow(t'')\rangle
\end{array}
\right).
\end{array}
\label{G}
\end{equation}
These Green's functions for the quantum dot system coupled to the normal and
superconducting leads have been obtained by the standard technique,
\cite{Yamada,Cuevas} 
which are explicitly represented in Appendix A. 

The sampling process must satisfy ergodicity and (as a
sufficient condition) detailed balance. For ergodicity, it is
enough to insert or remove the Ising variables with random
orientations at random times to generate all possible
configurations. To satisfy the detailed balance condition,
we decompose the transition probability as
\begin{eqnarray}
p(i\rightarrow j)&=& p^{prop}(i\rightarrow j)p^{acc}(i\rightarrow j)
\end{eqnarray}
where $p^{prop} (p^{acc})$ is the probability to propose (accept) the
transition from the configuration $i$ to the configuration $j$.
Here, we consider the insertion and removal of the Ising
spins as one step of the simulation process, which corresponds
to a change of $\pm 1$ in the perturbation order. The
probability of insertion/removal of an Ising spin is then
given by
\begin{eqnarray}
p^{prop}(n\rightarrow n+1)&=&\frac{1}{2}\frac{dt}{2t}\\
p^{prop}(n+1\rightarrow n)&=&\frac{1}{n+1}.
\end{eqnarray}
For this choice, the ratio of the acceptance probabilities
becomes
\begin{eqnarray}
\frac{p^{prop}(n\rightarrow n+1)}{p^{prop}(n+1\rightarrow n)}&=&
\pm i\frac{2K}{n+1}\frac{{\rm det} \left[{\hat N}^{(n)}\right]}
{{\rm det}\left[{\hat N}^{(n+1)}\right]},
\end{eqnarray}
where $+i (-i)$ corresponding to a spin which is inserted 
on the forward (backward) contour.

When the Metropolis algorithm is used to sample the
configurations, we accept the transition from $n$ to $n\pm 1$ with
the probability
\begin{eqnarray}
{\rm min}\left[1,\frac{p^{prop}(n\rightarrow n\pm 1)}{p^{prop}(n\pm 1\rightarrow n)}
\right].
\end{eqnarray}


In each Monte Carlo step, we can measure the Green function 
${\hat G}(t,t')$. By using Wick's theorem, the contribution of a certain
configuration is given by
\begin{equation}
\begin{array}{rcl}
\displaystyle
&&\frac{\displaystyle w_G\Big\{\left[(t_{k_1},s_{k_1}),\cdots,
(t_{k_n},s_{k_n})\right]; 
{\hat G}(t',t'')\Big\}}
{\displaystyle w\Big\{\left[(t_{k_1},s_{k_1}),\cdots,(t_{k_n},s_{k_n})\right]
\Big\}}\\
&=&
{\rm det} \left[{\hat N}^{(n)}\right]
{\rm det}\left(
\begin{array}{cc}
\left[{\hat N}^{(n)}\right]^{-1}&{\hat G}_0(t_{k_i},t'')\\
{\hat G}_0(t',t_{k_j})(\hat{\Gamma}-\hat{I})&{\hat G}_0(t',t'')
\end{array}
\right)\\
&=&\displaystyle
{\hat G}_0(t',t'')\nonumber\\
&+&\displaystyle i\sum_{ij}{\hat G}_0(t',t_{k_i})
\left[\left(\hat{\Gamma}-\hat{I}\right)\hat{N}
\right]_{ij}{\hat G}_0(t_{k,j},t'').
\end{array}
\end{equation}
The expectation values of the particle number $N(t)=\sum_\sigma\langle n_\sigma(t)\rangle$, pairing correlations 
$P(t)=\langle c_\uparrow(t) c_\downarrow(t)\rangle$, 
and double occupancy $D(t)=\langle n_\uparrow(t) n_\downarrow(t)\rangle$ 
at the quantum dot are calculated as
\begin{eqnarray}
N(t)&=&2-i\langle G_{11}(t,t)\rangle+i\langle G_{22}(t,t)\rangle,\\
P(t)&=&i\langle G_{12}(t,t)\rangle,\\
D(t)&=&1-N(t)+
\langle {\rm det} {\hat G}(t,t)\rangle.
\end{eqnarray}

We also measure the current from the quantum dot to the $\alpha$th lead,
which is given as
\begin{eqnarray}
I_\alpha&=&-2{\rm Im}\sum_{k\sigma}
V_{k\alpha\sigma}\langle c_{k\alpha\sigma}^\dag d_\sigma\rangle.
\end{eqnarray}
For convenient, we use the composite operator 
$\tilde{c}_{\alpha\sigma}=\sum_k V_{k\alpha\sigma}c_{k\alpha\sigma}$ and 
consider the following matrix as
\begin{equation}
\begin{array}{rcl}
{\hat A}_{0\alpha}(t'_k,t''_k)&=&\left\{
\begin{array}{ll}
{\hat A}_{0\alpha}^<(t',t''),& t'_k\le t''_k\\
{\hat A}_{0\alpha}^>(t',t''),& t'_k > t''_k
\end{array}
\right.,
\end{array}\label{A}
\end{equation}
where
\begin{equation}
\begin{array}{rcl}
{\hat A}_{0\alpha}^<(t',t'')&=&i\left(
\begin{array}{cc}
\langle \tilde{c}_{\alpha\uparrow}^\dag(t'')d_\uparrow(t')\rangle&
\langle \tilde{c}_{\alpha\downarrow}(t'')d_\uparrow(t')\rangle\\
\langle \tilde{c}_{\alpha\uparrow}^\dag(t'')d_\downarrow^\dag(t')\rangle&
\langle \tilde{c}_{\alpha\downarrow}(t'')d_\downarrow^\dag(t')\rangle
\end{array}
\right),\\
{\hat A}_{0\alpha}^>(t',t'')&=&-i\left(
\begin{array}{cc}
\langle d_\uparrow(t')\tilde{c}_{\alpha\uparrow}^\dag(t'')\rangle&
\langle d_\uparrow(t')\tilde{c}_{\alpha\downarrow}(t'')\rangle\\
\langle d_\downarrow^\dag(t')\tilde{c}_{\alpha\uparrow}^\dag(t'')\rangle&
\langle d_\downarrow^\dag(t')\tilde{c}_{\alpha\downarrow}(t'')\rangle
\end{array}
\right).
\end{array}
\end{equation}
The contribution for the matrix ${\hat A}_{0\alpha}(t',t'')$ 
of a certain configuration is given by
\begin{equation}
\begin{array}{rcl}
\displaystyle
&&\frac{\displaystyle w_A\Big\{\left[(t_{k_1},s_{k_1}),\cdots,
(t_{k_n},s_{k_n})\right]; 
{\hat A}_\alpha(t',t'')\Big\}}
{\displaystyle w\Big\{\left[(t_{k_1},s_{k_1}),\cdots,(t_{k_n},s_{k_n})\right]
\Big\}}\\
&=&\displaystyle 
{\rm det} \left[{\hat N}^{(n)}\right]
{\rm det}\left(
\begin{array}{cc}
\left[{\hat N}^{(n)}\right]^{-1}&{\hat A}_{0\alpha}(t_{k_i},t'')\\
{\hat G}_0(t',t_{k_j})(\hat{\Gamma}-\hat{I})&{\hat A}_{0\alpha}(t',t'')
\end{array}
\right)\\
&=&\displaystyle
{\hat A}_{0\alpha}(t',t'')\nonumber\\
&+&\displaystyle i\sum_{ij}{\hat G}_0(t',t_{k_i})
\left[\left(\hat{\Gamma}-\hat{I}\right)\hat{N}\right]_{ij}{\hat A}_{0\alpha}(t_{k_j},t'').
\end{array}
\end{equation}
Then we obtain the measurement formula for the currents as
\begin{eqnarray}
I_\alpha(t)&=&-2{\rm Im}\Big[ \langle {\hat A}_\alpha(t,t)\rangle_{11}
-\langle {\hat A}_\alpha(t,t)\rangle_{22}\Big].
\end{eqnarray}

This algorithm is essentially the same as 
the CTQMC method for the equilibrium state,\cite{KogaWerner} 
and thereby it is straightforward to modify the codes 
to deal with the nonequilibrium system.
Here, we comment on the dynamical sign problem: 
the weight for a certain configuration is, in general, represented
by the complex number [see eq. (\ref{weight})].
As discussed in the previous works,\cite{WernerOka,Werner} 
the dynamical sign problem becomes more serious 
in the simulations on longer contours and 
accurate measurements of physical quantities
are restricted to a certain time $t_{max}$.
The introduction of the coupling strength $\Gamma_S$ reduces the sign problem, 
which allows us to perform the simulations on longer contours.
On the other hand, the bias voltage $V$ little affects the dynamical sign problem.
Therefore, 
performing Monte Carlo simulations with a fixed $t_{max}$,
we can equally treat the systems with different values of $V$, where
the precision of the obtained results little varies. 
This is contrast to the perturbative approach, where more accurate results should be
obtained in the vicinity of $V=0$.

In this study, we use the coupling constant of the normal lead $\Gamma_N$ as
the unit of energy, and fix the superconducting gap as $\Delta/\Gamma_N=0.5$.
In the following, we perform the CTQMC simulations
to discuss nonequilibrium behavior at zero temperature
in the quantum dot coupled to 
the normal and superconducting leads.

\section{Results}\label{4}

In this section, we discuss how the interaction quench
affects the time evolution of the physical quantities.
Furthermore, by extrapolating them in the $t\rightarrow \infty$ limit,
we discuss steady-state properties of the quantum dot system.

\begin{figure}[htb]
\begin{center}
\includegraphics[width=7cm]{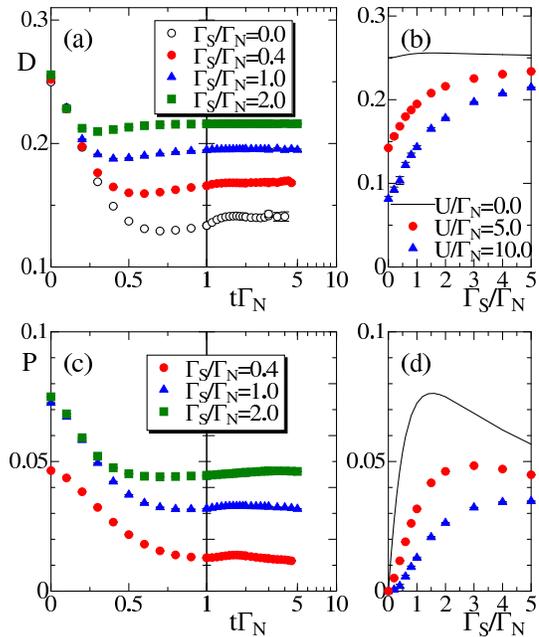}
\end{center}       
\caption{
The double occupancy (upper panels) and pairing correlation (lower panels) 
in the system with $T=0$, $V=0$, $\Delta/\Gamma_N=0.5$ and $\epsilon_d+U/2=0$.
The time evolutions are shown in (a) and (c), and 
the quantities at the steady state are shown in (b) and (d).
}
\label{fig:dble}
\end{figure}

First, we consider the quantum dot system without the bias voltage.
The time evolutions of the double occupancy and pair correlation 
are shown in Figs. \ref{fig:dble} (a) and (c).
In these figures, the quantities are shown on the linear plot 
in the initial relaxation region $(t\Gamma_N<1)$
and on the logarithmic plot in the rest $(t\Gamma_N>1)$.
When $\Gamma_S=0$, the quantum dot is only coupled to the normal lead 
and the system is reduced to the conventional Anderson impurity model,
where our results reproduce the previous ones.~\cite{WernerOka}
We find that the interaction quench decreases the double occupancy
($t\Gamma_N<1)$ and the system approaches the steady state $(t\Gamma_N>2)$.
When the superconducting lead couples to the quantum dot,
the double occupancy and pairing correlation for the initial state
$(t\le 0)$ increase due to the proximity effect.
As the interaction is turned on at $t=0$, the double occupancy
slightly decreases 
and the system quickly approaches the steady state, 
by comparing with the case $\Gamma_S=0$, as seen in Fig. \ref{fig:dble} (a).
This implies that electron correlations become less important in the system.
Although $t_{max}$ is finite, 
two quantities seem to converge around $t=t_{max}$.
This allows us to discuss the steady-state properties in the system.

Regarding the physical quantities at $t=t_{max}$ 
as those for the steady state, 
we discuss the ground state properties.
The results are shown in Figs. \ref{fig:dble} (b) and (d).
When the quantum dot is only contacted to the normal lead $(\Gamma_S=0)$,
the Coulomb interaction decreases the double occupancy
and the Kondo-singlet dominant state is realized.
As the coupling strength $\Gamma_S$ increases,
the double occupancy $D$ and pair correlation $P$ increase
due to the proximity effect. 
In the large $\Gamma_S$ region, 
the proximity-induced on-dot singlet-pairing dominant state is realized
and electron correlations become irrelevant.
In fact, the double occupancy approaches the value in the noninteracting limit 
$(D\rightarrow 1/4)$.
Therefore, the crossover occurs 
between the Kondo-singlet and proximity-induced singlet-pairing 
dominant states.\cite{Yamada}

\begin{figure}[htb]
\begin{center}
\includegraphics[width=7cm]{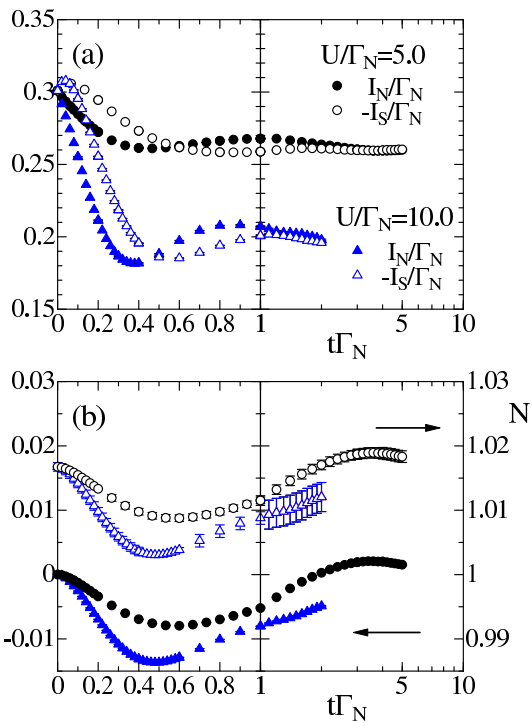}
\end{center}       
\caption{
The results are obtained by the CTQMC method in the quantum dot system 
with $\Gamma_S/\Gamma_N=1$, $V/\Gamma_N=0.5$, 
$\Delta/\Gamma_N=0.5$ and $T=0$ 
when $U/\Gamma_N=5.0$ (circles) and $U/\Gamma_N=10.0$ (triangles).
(a) Solid (open) symbols represent the time evolution of the current 
$I_N\;(-I_S)$. 
(b) Solid (open) symbols represent the integrated total currents 
$\sum_\alpha\int_0^t I_\alpha(t') dt'$ and particle number 
$N(t)$ at the quantum dot.
}
\label{fig:combine}
\end{figure}

When the bias voltage is applied, 
the current begins to flow between leads.
In the steady state $(t\rightarrow \infty)$, the current $I$ through the quantum dot
is constant ($I=I_N=-I_S$).
In the transient regime,
the current from the dot to the $\alpha$th lead is 
affected by the interaction $U$ and hybridization $\Gamma_\alpha$, 
which results in different values 
at a certain time $t$.
Therefore, we calculate both currents independently.
The results for $\Gamma_S/\Gamma_N=1$ are 
shown in Fig. \ref{fig:combine}.
In the initial state $(t\le 0)$, the currents $I_\alpha$ 
are given by the steady current through 
the noninteracting dot.
The introduction of the interaction decreases the currents $|I_N|$ and $|I_S|$
differently, and oscillation behavior appears, 
as shown in Fig. \ref{fig:combine} (a).
Finally, $I_N\sim -I_S$ and 
the system approaches the steady state.
We note that the steady current can be extrapolated since 
the relaxation of oscillation behavior appears in the transient current.

Here, we check the law of conservation of charge, 
which should be described as
\begin{eqnarray}
\sum_\alpha \int_0^t I_\alpha(t') dt' = 
N(t)-N(0).\label{cons}
\end{eqnarray}
Fig. \ref{fig:combine} (b) shows the integrated total currents 
from the quantum dot
[left-hand side of eq. (\ref{cons})] and 
particle number at the quantum dot $N(t)$.
It is found that the difference of two quantities is always constant,
which means that the conservation law, eq. (\ref{cons}), is satisfied 
within our numerical accuracy.
This relation provides a good check for the numerical implementation.

\begin{figure}[htb]
\begin{center}
\includegraphics[width=7cm]{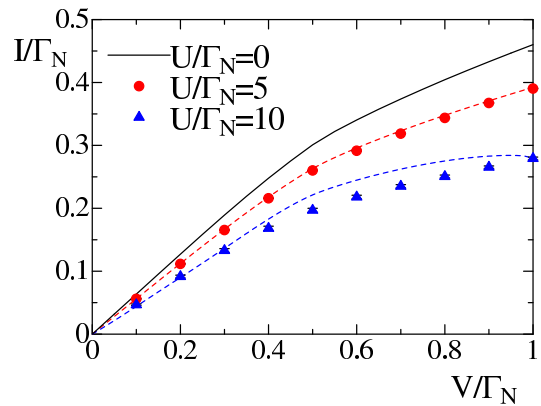}
\end{center}       
\caption{
Current-voltage characteristics in the system with 
$\Gamma_S/\Gamma_N=1$, $\Delta/\Gamma_N=0.5$, and $T=0$.
Circles and triangles represent the CTQMC results for $U/\Gamma_N=5.0$ and $10.0$ 
at $t=t_{max}$.
Dashed lines represent the MPT results 
at the low temperature $T=0.01\Gamma_N$.\cite{Yamada}
}
\label{fig:IV}
\end{figure}
Regarding the average of two currents at $t=t_{max}$
as the steady current $[I=(|I_N|+|I_S|)/2]$,
we obtain the current-voltage characteristics in the systems 
with $U/\Gamma_N=0.0, 5.0$, and $10.0$, as shown in Fig. \ref{fig:IV}.
As the error is defined as $\Delta I=|I_N(t_{max})-I|$,
it is smaller than the size of symbols in the figure.
When the bias voltage increases, the current monotonically increases,
together with the kink structure around $V=\Delta$.
We also find that the increase of the Coulomb repulsion decreases the current.
This implies that the interacting quantum dot can be regarded as 
the tunnel barrier on the interface.
Although the maximum time in our simulations is limited 
due to the dynamical sign problem 
($t_{max}\Gamma_N=5.0$ for $U/\Gamma_N=5.0$ and 
$t_{max}\Gamma_N=2.0$ for $U/\Gamma_N=10.0$),
the CTQMC method reproduces reasonable results.
In fact, in the weak coupling and small voltage region, 
our CTQMC data are in good agreement with the MPT results 
at a very low temperature 
$T/\Gamma_N=0.01$.~\cite{Yamada}
On the other hand, when $V\sim \Delta$, 
the CTQMC data are away from the other.
Since the precision of our data little depends on the bias voltage,
this may suggest that the MPT is not appropriate for quantitative discussions 
in this nonlinear response region $V\sim \Delta$ with $U/\Gamma_N\ge10$.

\begin{figure}[htb]
\begin{center}
\includegraphics[width=7cm]{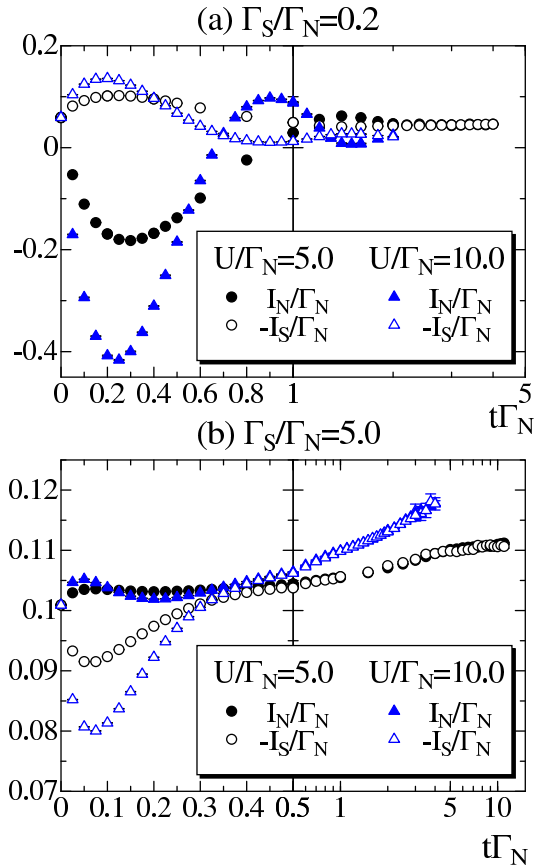}
\end{center}       
\caption{
(Color online) Time evolution of the currents for the systems with
$V/\Gamma_N=0.5$, $\Delta/\Gamma_N=0.5$, and $T=0$. 
Open (closed) symbols represent
the current from the superconducting (normal) lead to the quantum dot
when $\Gamma_S/\Gamma_N=0.2$ (a) and 
$\Gamma_S/\Gamma_N=5.0$ (b).
}
\label{fig:i2t}
\end{figure}

When the ratio $\Gamma_S/\Gamma_N$ is away from unity, 
interesting behavior appears.
The time evolutions of the currents for the systems with 
$\Gamma_S/\Gamma_N=0.2$ and $5.0$ are shown in Fig. \ref{fig:i2t}.
In the Kondo-singlet dominant region $(\Gamma_S/\Gamma_N=0.2)$, 
the interaction quench
leads to a drastic change in the current $I_N$,
in contrast to $I_S$, as shown in Fig. \ref{fig:i2t} (a).
When $t\Gamma_N\sim 0.02$, the current $I_N$ changes its sign and
a fairly large transient current flows against the bias voltage
around $t\Gamma_N\sim 0.3$.
This may be explained by the following.
In the initial steady state, the particle number at the quantum dot 
is larger than the half filling.
Therefore, the interaction quench tends to decrease the particle number
at the quantum dot, which results in the decrease of both currents from the dots.
In this case, there may be the relation 
$\Delta I_S/\Delta I_N\sim \Gamma_N/\Gamma_S$ around $t\sim 0$,
which induces a large transient current between the quantum dot and 
the normal lead.
As time progresses, the current $I_N$ turns over again
and the system approaches the steady state.
In the case $U/\Gamma_N=10.0$, 
the current largely fluctuates in the transient region $t\Gamma_N<1$ and 
the system may not reach the steady state around $t=t_{max}$.
However, its oscillation rapidly relaxes and 
the steady current is expected to be around two currents at $t=t_{max}$.

On the other hand, when the superconducting lead tightly 
couples to the quantum 
dot ($\Gamma_S/\Gamma_N=5.0$), 
the currents are slightly changed by the interaction quench
and the magnitudes of two currents approach each other 
around $t\Gamma_N\sim 0.5$, 
in contrast to the $\Gamma_S/\Gamma_N=0.2$ case.
Nevertheless, we do not find the convergence in the current 
even around $t=t_{max}$,
as shown in Fig. \ref{fig:i2t}.
This may originate from the time evolution of the Kondo resonance
induced by the interaction quench.
It is characterized by an exponentially small energy scale and
should be slow to be built up in the case $\Gamma_S/\Gamma_N>1$. 
In this case, the steady current, as which 
the transient currents at $t=t_{max}$ are regarded in the paper, 
should be a lower bound of the correct value.
Therefore, Monte Carlo simulations on longer contours are necessary 
to obtain the steady current more precisely.\cite{Werner}

\begin{figure}[htb]
\begin{center}
\includegraphics[width=7cm]{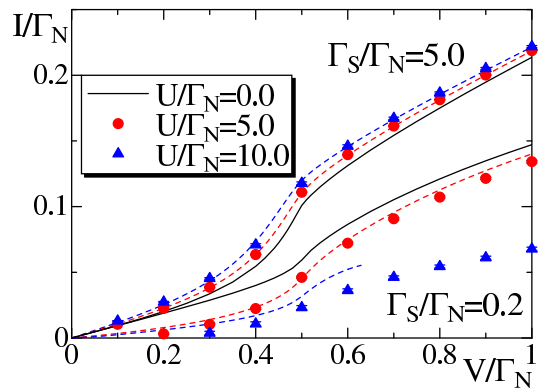}
\end{center}       
\caption{
Current-voltage characteristics in the system with 
$\Gamma_S/\Gamma_N=0.2$ and $5.0$ when
$\Delta/\Gamma_N=0.5$, $\epsilon_d+U/2=0$ and $T=0$.
Circles and triangles represent the CTQMC results for 
$U/\Gamma_N=5.0$ and $10.0$ 
at $t=t_{max}$.
Dashed lines represent the MPT results 
at the low temperature $T=0.01\Gamma_N$.\cite{Yamada}
}
\label{fig:IV2}
\end{figure}

By performing similar calculations, 
we obtain the current-voltage characteristics of the systems with 
$\Gamma_S/\Gamma_N=0.2$ and $5.0$, as shown in Fig. \ref{fig:IV2}.
In the noninteracting case $(U=0)$,
the system should be regarded as the normal-superconducting junction 
with a simple tunnel barrier.
When $\Gamma_S/\Gamma_N$ is away from unity,
the current rapidly increases around $V\sim \Delta$.
This behavior is well described by the conventional BTK theory.\cite{BTK}
The introduction of the Coulomb interaction 
increases the current through the quantum dot tightly coupled 
to the superconducting lead $(\Gamma_S/\Gamma_N=5.0)$.
In the case, we could not obtain the steady current precisely,
as discussed above. 
Nevertheless, our results are consistent with 
those obtained by the MPT,\cite{Yamada}
which should be appropriate in this region.
On the other hand, 
in the Kondo-singlet dominant region $(\Gamma_S/\Gamma_N=0.2)$, 
the Coulomb interaction decreases the current.
In the MPT, the correlation effects are underestimated
and the difference between two results becomes large 
in the strong coupling region.

\begin{figure}[htb]
\begin{center}
\includegraphics[width=7cm]{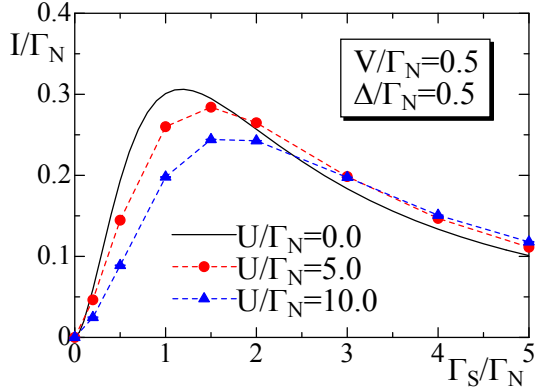}
\end{center}       
\caption{
Currents functions of $\Gamma_S/\Gamma_N$ 
in the system with $V/\Gamma_N=0.5$, 
$\Delta/\Gamma_N=0.5$, $\epsilon_d+U/2=0$ and $T=0$.
}
\label{fig:IGS}
\end{figure}

We finally discuss the effect of electron correlations 
in the system with a fixed voltage $V/\Gamma_N=0.5$. 
The results for $U/\Gamma_N=0.0, 5.0$, and $10.0$ are shown in Fig. \ref{fig:IGS}.
The increase in $U$ monotonically decreases (increases)
the current when $\Gamma_S/\Gamma_N<1.5 (\Gamma_S/\Gamma_N>4)$.
In the intermediate region $(\Gamma_S/\Gamma_N\sim 2.5)$, 
nonmonotonic behavior appears:
the introduction of the interaction once increases the current in 
the proximity-induced on-dot pairing-dominant region. 
On the other hand, a further increase of the interaction decreases the current,
where the Kondo singlet becomes dominant.
The crossover behavior in the nonlinear response region 
may be similar to that in the linear response region of
the quantum dot system with $\Delta\rightarrow \infty$,\cite{Tanaka}
where the zero bias conductance has a maximum around $\Gamma_S\sim U/2$.
Therefore, we can say that the crossover between the Kondo-singlet and 
proximity-induced on-dot pairing dominant states affects 
the current-voltage characteristics under the high voltage beyond 
the superconducting gap $V\ge\Delta$.

In this paper, 
we have discussed the nonequilibrium phenomena of
the quantum dot coupled to the normal and superconducting leads
by means of the CTQMC method in the Keldysh formalism.
It is an interesting problem how the local Coulomb interaction affects 
the Josephson current and multiple Andreev reflections
in the quantum dot coupled to two superconducting leads,
which is now under consideration.

\section{Summary}\label{5}
We have quantitatively studied the nonequilibrium phenomena through the quantum dot
coupled to the normal and superconducting leads,
extending the weak-coupling CTQMC method to treat 
the superconducting state in the Nambu formalism.
We have confirmed that the obtained results are in good agreement 
with those obtained by the MPT
in the weak coupling region and $\Gamma_S/\Gamma_N>1$.\cite{Yamada}
In the strong coupling region,
we have clarified that the crossover between the Kondo singlet dominant 
and Cooper pairing singlet dominant regions affects 
the current-voltage characteristics.
Transient currents induced by the interaction quench have been discussed.

\section{Acknowledgments}
We would like to thank R. Sakano, Y. Yamada, and P. Werner
for valuable discussions.
We also thank Y. Yamada for providing data, 
which we used in Figs. \ref{fig:IV} and \ref{fig:IV2}.
This work was partly supported by the Global
COE Program ``Nanoscience and Quantum Physics" from
the Ministry of Education, Culture, Sports, Science and
Technology (MEXT) of Japan.
A part of computations was carried out on TSUBAME2.0 at Global Scientific 
Information and Computing Center of Tokyo Institute of Technology and 
on the Supercomputer Center at the 
Institute for Solid State Physics, University of Tokyo.
The simulations have been performed using some of the ALPS libraries.
\cite{alps1.3}

\section{Appendix}
\subsection{Calculation of noninteracting Green's functions}

Here, we explicitly show the impurity Green's functions 
in the noninteracting case $U=0$.
The Green's functions for the impurity (quantum dot) site are represented 
in terms of the hybridization functions ${\hat \Delta}(\omega)$ as,
\begin{eqnarray}
{\hat G}_0^{R,A}(\omega)&=&\left[(\omega\pm i\eta){\hat \sigma}_0
-\epsilon_d{\hat \sigma}_z-{\hat\Delta}^{R,A}(\omega)\right]^{-1}\\
{\hat G}_0^{</>}(\omega)&=&{\hat G}_0^R(\omega){\hat\Delta}^{</>}(\omega)
{\hat G}_0^A(\omega),
\end{eqnarray}
where the hybridization functions are given as
\begin{eqnarray}
{\hat\Delta}^{R,A}(\omega)&=&\sum_{\alpha=N,S}
{\hat\Delta}^{R,A}_\alpha(\omega)\\
{\hat\Delta}^R_N(\omega)&=&-i\Gamma_N{\hat \sigma}_0\\
{\hat\Delta}^R_S(\omega)&=&-i\Gamma_S\beta(\omega){\hat M}(\omega)\\
{\hat\Delta}^A_\alpha(\omega)&=&\left[{\hat\Delta}^R_\alpha(\omega)\right]^+\\
{\hat\Delta}^<_N(\omega)&=&2i\Gamma_N {\hat F}(\omega)\\
{\hat\Delta}^<_S(\omega)&=&2i\Gamma_S {\rm Re}\left[\beta(\omega)\right]
f(\omega){\hat M}(\omega)\\
{\hat\Delta}^>_N(\omega)&=&-2i\Gamma_N \left({\hat\sigma}_0-{\hat F}(\omega)\right)\\
{\hat\Delta}^>_S(\omega)&=&-2i\Gamma_S {\rm Re}\left[\beta(\omega)\right]
\left[1-f(\omega)\right]{\hat M}(\omega),
\end{eqnarray}
where
${\hat F}(\omega)={\rm diag}\left[ f(\omega-\mu_N), f(\omega+\mu_N)\right],
{\hat M}(\omega)={\hat\sigma}_0-\frac{\Delta_S}{\omega}{\hat\sigma}_x$ and 
$\beta(\omega)=\frac{|\omega|}{\sqrt{\omega^2-\Delta^2}}\theta(|\omega|-\Delta)
-\frac{i\omega}{\sqrt{\Delta^2-\omega^2}}\theta(\Delta-|\omega|)$.
Using the Fourier transformations, 
we obtain the lesser and greater Green's functions eq. (\ref{G}),
\begin{eqnarray}
{\hat G}_0^{</>}(t)&=&\int \frac{d\omega}{2\pi}
{\hat G}_0^{</>}(\omega)e^{-i\omega t}.
\end{eqnarray}

\subsection{Calculation of ${\hat A}_{0\alpha}(t,t')$}

We obtain the expression of the matrix ${\hat A}_{0\alpha}(t,t')$,
which is necessary to calculate the current from the quantum dot 
to the $\alpha$th lead.
We use the expression in previous papers\cite{WernerOka,Jauho}
as
\begin{equation}
\left(
\begin{array}{cc}
{\hat A}_{0\alpha}^R&{\hat A}_{0\alpha}^K\\
0&{\hat A}_{0\alpha}^A 
\end{array}
\right)=-i\left(
\begin{array}{cc}
{\hat G}_{dd}^R&{\hat G}_{dd}^K\\
0&{\hat G}_{dd}^A 
\end{array}
\right)\left(
\begin{array}{cc}
{\hat \Delta}_\alpha^R&{\hat \Delta}_\alpha^K\\
0&{\hat \Delta}_\alpha^A 
\end{array}
\right),
\end{equation}
where ${\hat G}_{dd}^{\eta}(\eta=R, A, K)$ is the retarded, advanced, and Keldysh 
impurity Green's function for the Anderson model
eq. (\ref{Hami}) and ${\hat \Delta}^K_\alpha={\hat \Delta}^<_\alpha+{\hat \Delta}^>_\alpha$.
Using the Fourier transformations as 
\begin{eqnarray}
{\hat A}_{0\alpha}^{</>}(t)&=&\int \frac{d\omega}{2\pi}
{\hat A}_{0\alpha}^{</>}(\omega) e^{-i\omega t},\\
{\hat A}_{0\alpha}^{</>}(\omega)&=&\frac{1}{2}
\left[{\hat A}_{0\alpha}^K(\omega)\mp{\hat A}_{0\alpha}^R(\omega)\pm 
{\hat A}_{0\alpha}^A(\omega)\right],
\end{eqnarray}
we obtain the matrix ${\hat A}_{0\alpha}(t,t')$ in eq. (\ref{A}).

%


\end{document}